

\font\twrm=cmr10 scaled\magstep1
\font\twi=cmmi10 scaled\magstep1
\font\twsy=cmsy10 scaled\magstep1
\font\twit=cmti10 scaled\magstep1
\font\twbf=cmbx10 scaled\magstep1
\font\twsrm=cmr7 scaled\magstep1
\font\twsi=cmmi7 scaled\magstep1
\font\twssy=cmsy7 scaled\magstep1
\font\twsbf=cmbx7 scaled\magstep1
\font\twfrm=cmr5 scaled\magstep1
\font\twfi=cmmi5 scaled\magstep1
\font\twfsy=cmsy5 scaled\magstep1
\font\twfbf=cmbx5 scaled\magstep1

\font\nrm=cmr9
\font\ni=cmmi9
\font\nsy=cmsy9
\font\nit=cmti9
\font\nbf=cmbx9


\catcode`@=11 
\def\twelve{\def\rm{\fam\z@\twrm}
  \textfont\z@=\twrm \scriptfont\z@=\twsrm \scriptscriptfont\z@=\twfrm
  \textfont\@ne=\twi \scriptfont\@ne=\twsi \scriptscriptfont\@ne=\twfi
  \textfont\tw@=\twsy \scriptfont\tw@=\twssy \scriptscriptfont\tw@=\twfsy
  \textfont\itfam=\twit  \def\it{\fam\itfam\twit }
  \textfont\bffam=\twbf \scriptfont\bffam=\twsbf \scriptscriptfont\bffam=\twfbf
  \def\bf{\fam\bffam\twbf }
  \normalbaselineskip=14pt plus .1pt
  \normalbaselineskip\rm}
\def\ten{\def\rm{\fam\z@\tenrm}
  \textfont\z@=\tenrm \scriptfont\z@=\sevenrm \scriptscriptfont\z@=\fiverm
  \textfont\@ne=\teni \scriptfont\@ne=\seveni \scriptscriptfont\@ne=\fivei
  \textfont\tw@=\tensy \scriptfont\tw@=\sevensy \scriptscriptfont\tw@=\fivesy
  \textfont\itfam=\tenit  \def\it{\fam\itfam\tenit }
  \textfont\bffam=\tenbf \scriptfont\bffam=\sevenbf
  \scriptscriptfont\bffam=\fivebf \def\bf{\fam\bffam\tenbf }
  \normalbaselineskip=12pt plus .1pt
  \normalbaselineskip\rm}
\def\nine{\def\rm{\fam\z@\nrm}
  \textfont\z@=\nrm \scriptfont\z@=\sevenrm \scriptscriptfont\z@=\fiverm
  \textfont\@ne=\ni \scriptfont\@ne=\seveni \scriptscriptfont\@ne=\fivei
  \textfont\tw@=\nsy \scriptfont\tw@=\sevensy \scriptscriptfont\tw@=\fivesy
  \textfont\itfam=\nit  \def\it{\fam\itfam\nit }
  \textfont\bffam=\nbf \scriptfont\bffam=\sevenbf
  \scriptscriptfont\bffam=\fivebf   \def\bf{\fam\bffam\nbf }
  \normalbaselineskip=11pt plus .1pt
  \normalbaselineskip\rm}
\catcode`@=12

\def\today{\number\year\space \ifcase\month\or 	January\or February\or
	March\or April\or May\or June\or July\or August\or September\or
	October\or November\or December\fi\space \number\day}


\def\degrees{\hbox{${}^\circ$\hskip-3pt .}}

\def\aref#1#2;#3;#4;#5;#6.{\itemitem{#1} #2 {\it #4} {\bf #5} (#3) #6.}
\def\abook#1#2;#3;#4;#5;#6;#7.{\itemitem{#1} #2, in {\it #4}, #5 (#6, #3),
                                                                       #7.}
\def\apress#1#2;#3;#4;#5.{\itemitem{#1} #2, {\it #4} (#3), #5.}
\def\arep#1#2;#3;#4.{\itemitem{#1} #2 (#3) {#4}.}

\def\spose#1{\hbox to 0pt{#1\hss}}
\def\simlt{\mathrel{\spose{\lower 3pt\hbox{$\mathchar"218$}}
     \raise 2.0pt\hbox{$\mathchar"13C$}}}
\def\simgt{\mathrel{\spose{\lower 3pt\hbox{$\mathchar"218$}}
     \raise 2.0pt\hbox{$\mathchar"13E$}}}
\def\trms{T_{\rm rms}}
\def\eq{Eq.$\,$}
\def\frac#1/#2{\leavevmode\kern.1em
 \raise.5ex\hbox{\the\scriptfont0 #1}\kern-.1em
 /\kern-.15em\lower.25ex\hbox{\the\scriptfont0 #2}}

\headline={\ifnum\pageno=1\firstheadline\else
\ifodd\pageno\rightheadline \else\leftheadline\fi\fi}
\def\firstheadline{\hfil}
\def\rightheadline{\hfil}
\def\leftheadline{\hfil}
	\footline={\ifnum\pageno=1\firstfootline\else\otherfootline\fi
}
\def\firstfootline{\rm\hss\folio\hss}
\def\otherfootline{\hfil}

\parindent=1.5pc
\hsize=6.0truein
\vsize=8.5truein
\nopagenumbers


\line{\hfil CfPA-94-TH-28}
\line{\hfil astro-ph/9406060}
\bigskip

\ten
\centerline{\bf QUOTING EXPERIMENTAL INFORMATION}
\baselineskip=16pt
\vglue 0.8cm
\centerline{MARTIN WHITE \& DOUGLAS SCOTT}
\baselineskip=13pt
\centerline{\it Center for Particle Astrophysics, University of California,}
\baselineskip=12pt
\centerline{\it Berkeley, CA 94720-7304}
\vglue 0.8cm
\centerline{ABSTRACT}
\vglue 0.3cm
{\rightskip=3pc
 \leftskip=3pc
 \baselineskip=12pt\noindent
There was a question raised at this meeting as to the best way to present the
results of experiments measuring anisotropies in the CMB.  Here we will
make some simple comments about the 3 main competing methods and some
suggestions.  In particular we will give an over-simplified but hopefully
useful method of comparing GACF numbers with other power spectra.
\vglue 0.3cm
\noindent{\it Subject headings:} cosmic background radiation ---
cosmology: theories and observations --- GACF's: justified
\vglue 0.6cm}

\vfil
\twelve
\baselineskip=14pt

\vglue 0.6cm
\leftline{\bf 1.~Introduction}
\vglue 0.4cm
There are a number of ways of quoting the level of anisotropy measured by
a CMB experiment.  Some of the most common are:
the rms temperature measured by the experiment $\trms$;
the amplitude and correlation angle of a Gaussian AutoCorrelation Function
(GACF) $C_0$ and $\theta_c$;
the amplitude of a `flat' or Harrison-Zel'dovich power spectrum;
and the amplitude of a `standard' spectrum such as CDM.

The rms temperature measured by the experiment is the quantity
most directly related to the
data, but also the least informative when it comes to comparing different
experiments, since it depends on the window function and calibration procedures
etc.~of the experiments.
It also does not include the sample${}^1$ and cosmic variance associated with
the measurement.
All the other methods choose to quote the amplitude of fluctuations by assuming
the power spectrum has a certain form and fitting for the amplitude.
This quantity is independent of the window function normalization, calibration
etc., and if the amplitude really is fit properly to the data, cosmic and
sample
variance are included in the result automatically.
As long as the {\it same} form is assumed for {\it all}
experiments, such numbers can be safely compared from experiment to experiment
and with theories.

Since the ability to compare experiments
is certainly something we would all like to have, it seems then
that fitting a power spectrum is the way to go.  One suggestion
(Scott Dodelson, this meeting), is to define a `standard' model such as CDM
with a fixed $\Omega_B$ and $h$, and to quote the amplitude of such a spectrum.
This has the advantage that, to the extent that such a model correctly
describes the fluctuations on the sky, the amplitudes so obtained can be
trivially compared from experiment to experiment: they should be all the same.
The drawbacks of this particular form for the power spectrum are that it
involves an enormous amount of theoretical prejudice at the level even of
{\it quoting} the experimental measurements, and that the `standard' power
spectrum has a large amount of structure in it.
Should the model turn out to be something other than `standard CDM' or if
$\Omega_B$ or $h$ turn out to be different than was assumed, this structure
has to be deconvolved from the measurements before a new power spectrum is
used.
We note in passing that comparison of $C_\ell$'s from different codes has not
been done in detail to let us understand the reproducability of
the curves.

At the other extreme are the flat spectrum and the GACF.  These are not at all
well motivated as models of our sky, but they are extremely simple power
spectra.  While measurements on different scales are not expected to give the
{\it same} amplitude (Doppler peaks give a larger amplitude for
experiments on degree scales, for example), it is not difficult to remember
a few numbers at a range of scales for comparison, and they have even been
tabulated for CDM${}^2$.

So what are the merits and disadvantages of GACF's and flat spectra?
Historically experimental results (upper limits until recently)
have been quoted in terms of the amplitude
of a `most sensitive' GACF.  This method pre-dates the emergence of window
functions into the popular consciousness and gives an alternate view of the
`scale' at which the experiment is sensitive and the degree of
the sensitivity${}^{3,4}$.
As discussed in references 2 and 5 the GACF analysis is straightforward to
understand in terms of window functions, and a conversion from GACF numbers to
`flat spectrum' numbers is easy to do${}^{2,6,7,8}$.
Note that the advantage of having a readily understandable scale in your fit
($\theta_c$) is also a {\it disadvantage} when it comes to comparing
experiments, since values of $C_0$ for different (arbitrary) $\theta_c$ cannot
be directly compared.
The flat spectrum has no such problem since it is featureless and easily
understandable, but does not directly contain the information on the window
function that the GACF encodes.

Here we would like to make some basic remarks about the relation of the GACF
to the window function and about converting from GACF's to flat spectra
and back.  The details of such a conversion have been treated before,
but we would like to present a simple (and not necessarily very accurate!)
method which hopefully illustrates the general idea.

\vglue 0.6truecm
\leftline{\bf 2. To and from the GACF}
\vglue 0.4truecm
Since our aim here is simplicity rather than rigour, we are going to assume
that
all CMB experiments on small scales are two-beam square-wave chop experiments,
with peak-to-peak chop $\alpha$ and (gaussian) beam width $\sigma$.
Under this approximation the window function is of the well known
form
$$\eqalign{
W_\ell &= 2\left[ 1-P_\ell(\cos\alpha) \right] e^{-\ell(\ell+1)\sigma^2} \cr
       &\simeq {\scriptstyle {1\over2}}
               (\ell\alpha)^2  e^{-(\ell\sigma)^2}\!, \cr } \eqno(1)
$$
where in the second line we have made the approximation $\alpha\ll1$, or more
correctly that $\ell\alpha<1$ for the $\ell$ range of interest.  This is not
a tremendously good approximation, but brings out the essential features of
what is going on.

Now in the language of power spectra and window functions, the most important
quantity measured by an experiment is the power through the window,
which is given by
$$
{\rm Power} = {1\over 4\pi}\sum_{\ell=2}^{\infty} (2\ell+1) C_\ell W_\ell.
\eqno(2)
$$
As experiments become more sensitive, other quantities (e.g.~correlations) are
going to play an important role${}^9$, but for now the power is what most
experiments are sensitive to${}^5$.
For experiments on small or intermediate angular scales we can approximate
the sum as an integral and replace $(2\ell+1)\rightarrow 2\ell$.  If the
window function drops off at small $\ell$ we are safe in extending the
lower limit of the integration to $\ell=0$.
The quantity normally plotted with the window function is
$\ell(\ell+1)C_\ell\approx\ell^2 C_\ell$ in terms of which
$$
{\rm Power}\approx {2\over 4\pi} \int_{-\infty}^\infty  \ell^2C_\ell\ W_\ell
\ d\log\ell,
\eqno(3)
$$
which explains the logarithmic $\ell$-axis normally used in such plots.

What do the two `competing' power spectra look like?  The flat spectrum is
extremely simple
$$
\ell(\ell+1)C_\ell = {24\pi\over 5} Q_{\rm flat}^2 = {\rm constant}.
\eqno(4)
$$
The factor of $24\pi/5$ is there simply to soak up the constant in \eq(2)
so that $Q_{\rm flat}^2$ is the coefficient of $W_2$ {\'a} la COBE.
The GACF power spectrum on the other hand is
$$
\ell(\ell+1)C_\ell = 2\pi C_0\ \ell^2 \theta_c^2\, e^{-1/2\, \ell^2\theta_c^2}.
\eqno(5)
$$
Notice now the similarity between $\ell(\ell+1)C_\ell$ for the GACF and the
expression for the window function of \eq(1).
Keeping in mind that the power measured by the experiment is the fixed
quantity defined by the data, it will come as no surprise
that the $\theta_c$ which gives the lowest $C_0$ is
$\theta_c=\sqrt{2}\sigma\simeq0.6\times$FWHM.  This is the value of
$\theta_c$ which maximizes the power through the window function for a
GACF [using \eq(1) and \eq(5) in \eq(3)].
All that varying $\theta_c$ in the fit is doing is matching the power spectrum
to the window function of the experiment, and this is another way of encoding
the information about which $\ell$'s the experiment measures.
In the {\it highly simplified} approximation for window functions given
by \eq(1), the correlation angle of the experiment is set by the beam size,
which also sets the peak of the window function, which is $\ell_0=1/\sigma$.
In practice this last relation is very poor (see Table~1), since getting the
peak correct requires accurately modelling the left hand rise of $W_\ell$ well
for `large' $\ell\alpha$ unless, $\sigma\gg\alpha$.  This needs a detailed
consideration of each experiment${}^{10}$.

We can do some easy integrals to see how the GACF-to-flat spectrum
conversion works in our over-simplified but hopefully instructive
approximation.  The power through the window of the flat spectrum is
proportional to the area under the window function vs.~$\log\ell$, or
$$\eqalign{
{\rm Power} &= {6\over 5}Q_{\rm flat}^2
               \int_0^\infty d\ell^2\ \ell^{-2}\,W_\ell\cr
&= {6\over 5}Q_{\rm flat}^2\ {\alpha^2\over\sigma^2}. \cr }
\eqno(6)
$$
While the power through the window for a GACF is
$$\eqalign{
{\rm Power} &= {1\over 2}C_0\theta_c^2
               \int_0^\infty d\ell^2\ e^{-1/2\ \ell^2\theta_c^2}\,W_\ell \cr
&= C_0\ {\alpha^2\over 4\sigma^2}, \cr }
\eqno(7)
$$
where in the second line we have evaluated the integral for
$\theta_c=\sqrt{2}\sigma$.
Relating these two we see that $Q_{\rm flat}\simeq 0.46C_0^{1/2}$,
which compares well with numbers obtained using the accurate $W_\ell$'s
and the $\theta_c$'s quoted for each experiment (see Table~1).
\vglue 0.6truecm
{\baselineskip=12pt\ten
\noindent{\bf Table~1:}
Summary of the peak of the window function and the $C_0^{1/2}$ to
$Q_{\rm flat}$ conversion factor for current CMB experiments on small and
intermediate scales.
Note that the peak of the window function does not correspond very well to
$1/\sigma$, showing that $\sigma\gg\alpha$ is not a good approximation for
most experiments.  However, the $C_0^{1/2}$ to $Q_{\rm flat}$ conversion is
close to $0.46$--$0.50$, for many experiments, as predicted by simple
calculation.}
\vglue 0.4truecm
\centerline{
\vbox{ \offinterlineskip
\halign{
\strut#&\vrule#&
#\hfil&\vrule#&
\hfil#\hfil&\vrule#&
\hfil#\hfil&\vrule#&
\hfil#\hfil&\vrule#&
\hfil#\hfil&\vrule#&
\hfil#\hfil&\vrule#\cr
\noalign{\hrule}
&&\ Experiment\ &&\quad$\ell_0$\quad&&\quad$1/\sigma$\quad&
&\quad$\theta_c$\quad&&\quad$Q/C_{0_{\vphantom{1}}}^{1/2}$\quad&\cr
\noalign{\hrule}
&&\ Tenerife &&   20${}^{\vphantom{1}}$&&   25&&  $4\degrees0$&& 0.50&\cr
&&\ SP91     &&   66&&   96&&  $1\degrees5$&& 0.44&\cr
&&\ SK93     &&   71&&   93&&  $1\degrees2$&& 0.49&\cr
&&\ Python   &&   73&&  180&&  $1\degrees0$&& 0.47&\cr
&&\ ARGO     &&  107&&  156&&  $0\degrees5$&& 0.42&\cr
&&\ MAX      &&  158&&  270&&  $0\degrees5$&& 0.45&\cr
&&\ MSAM2    &&  143&&  289&&  $0\degrees5$&& 0.44&\cr
&&\ MSAM3    &&  249&&  289&&  $0\degrees3$&& 0.50&\cr
\noalign{\hrule} }} }

\vglue 0.6truecm
We can do a little better for some experiments by noticing that they perform
a double-difference (`triple-beam') rather than a single difference
(`double-beam') measurement.
This means that the window function is better represented by the functional
form $W_\ell\propto\ell^4e^{-\ell(\ell+1)\sigma^2}$ than by the lower order
expression in \eq(1).  This changes the peak to $\ell_0=\sqrt{2}/\sigma$, and
the power of a GACF through this window peaks at $\theta_c=\sigma$.
The comparison of a GACF and a flat spectrum then leads to
$Q_{\rm flat}/C_0^{1/2}=2\sqrt{5}/9\simeq0.50$.
We would expect Tenerife, SK93, Python and MSAM3 to be better characterized
by this double-difference approximation.  A glance at Table~1 shows that the
$Q/C_0^{1/2}$ ratios are indeed higher.

\vglue 0.6truecm
\leftline{\bf 3. Conclusions}
\vglue 0.4truecm
In reality the approximations made above are not quite adequate for
accurately converting from GACF numbers to other measures of the power,
but the procedure has the virtue of being very straightforward:
one calculates the power in \eq(2) for whatever power spectrum was used in
the quoted answer.  Assuming that this is the important information that the
experiment provides, you adjust the normalization of your favourite power
spectrum to match this number.  Note that as long as you use the same window
function in both calculations, the $W_\ell$ normalization doesn't matter.
This is a boon because getting the shape of the window function is usually
a lot easier than determining the height!
Scaling from numbers quoted for power spectrum fits has the advantage that
{\it all} of the experimental and theoretical errors will be included
in such a fit.

Hopefully working through this over-simplified example has given some
insight into the relation between the well loved GACF method and the
$\ell$-space measures preferred by theorists.
\smallskip
\noindent{\it The bottom line}: If you want a quick and
dirty $C_0$ to $Q$ conversion, then a factor of two is good at the $10\%$
level for all experiments to date.
\vglue 0.6truecm
\noindent {\bf Acknowledgements}
\vglue 0.4truecm
We would like to thank Lawrence Krauss, Pete Kernan and all the organizers
for a successful conference and Mark Srednicki for many GACF-related
discussions/arguments.
M.W.~acknowledges the support of an SSC fellowship from the TNRLC.  This
work was also supported by grants from the NSF.
\vglue 0.6truecm
\leftline{\bf References}
\vglue 0.4truecm
\frenchspacing
\aref{1.} D. Scott, M. Srednicki and M. White;1994;ApJ;421;L5.
\apress{2.} M. White, D. Scott and J Silk;1994;Ann. Rev. A\&A;in press.
\abook{3.} A. N. Lasenby and R. D. Davies;1988;Large-Scale
Motions in the Universe;ed. V. C. Rubin and G. V. Coyne;Vatican Press
and Princeton;p.$\,397$.
\aref{4.} A. C. S. Readhead, C. R. Lawrence, S. T. Myers, W. L. W. Sargent,
H. E. Hardebeck and A. T. Moffet;1989;ApJ;346;566.
\apress{5.} E. F. Bunn, M. White, M. Srednicki and D. Scott;1994;ApJ;in press.
\apress{6.} J. R. Bond;1994;Astrophs. Lett. \& Commun. (Proceedings of the
Capri workshop);in press.
\arep{7.} D. Scott and M. White;1994;this volume.
\arep{8.} P. Steinhardt;1994;this volume.
\apress{9.} M. White;1994;A\&A;in press.
\aref{10.} M. Srednicki, M. White, D. Scott and E. F. Bunn;1993;Phys. Rev.
Lett.;71;3747.
\apress{11.} M. White and M. Srednicki;1994;Window Functions for CMB
Experiments;preprint CfPA-94-TH-11, astro-ph/9402037.
\bye